\documentclass[format=acmsmall, review=false, screen=true]{acmart}

\usepackage{booktabs} 

\begin{document}
\title[Intelligent Tutoring Systems]{Intelligent Tutoring Systems: A Comprehensive Historical Survey with Recent Developments}

\author{Ali Alkhatlan}
\affiliation{%
  \institution{University of Colorado Colorado Springs}
  \department[0]{College of Engineering and Applied Sciences}
  \department[1]{Department of Computer Science}
  \streetaddress{1420 Austin Bluffs Pkwy}
  \city{Colorado Springs}
  \state{Colorado}
  \postcode{80918}
  \country{USA}}
\email{aalkhatl@uccs.edu}

\author{Jugal K. Kalita}
\affiliation{%
  \institution{University of Colorado Colorado Springs}
  \department[0]{College of Engineering and Applied Sciences}
  \department[1]{Department of Computer Science}
  \streetaddress{1420 Austin Bluffs Pkwy}
  \city{Colorado Springs}
  \state{Colorado}
  \postcode{80918}
  \country{USA}}
\email{jkalita@uccs.edu}

\begin{abstract}

This paper provides interested beginners with an updated and detailed  introduction to the field of Intelligent Tutoring Systems (ITS). ITSs are computer programs that use artificial intelligence techniques to enhance and personalize automation in teaching. This paper is a literature review that provides the following: First, a review of the history of ITS along with a discussion on the interface between human learning and computer tutors and how effective ITSs are in contemporary education. Second, the traditional architectural components of an ITS and their functions are discussed along with approaches taken by various ITSs. Finally,   recent innovative ideas in ITS systems are presented. This paper concludes with some of the author's views regarding future work in the field of intelligent tutoring systems.

\end{abstract}

%
%

%
%


\maketitle

\renewcommand{\shortauthors}{Alkhatlan and Kalita}


\section{Introduction}

From the earliest days, computers have been employed in a variety of  areas to help society, including education. The computer was first introduced to the field of education in the 1970s under the aegis of Computer Assisted Instruction (CAI). Efforts at using computers in education were presented by Carbonel in the 1970s. He claimed that a CAI could be endowed with enhanced capabilities  by incorporating Artificial Intelligence (AI) techniques to overcome current limitations \cite{c1}.

In 1984, a study conducted by Bloom \cite{c141} showed that learners who studied a topic under the guidance of a human tutor, combined with traditional assessment and corrective instructions performed two standard deviations (sigma) better than those who received traditional group teaching. Researchers in the field of AI saw a solid opportunity to create intelligent systems to provide effective tutoring for  individual students, tailored to their needs and to enhance learning  \cite{c3}. Researchers found a new and inspiring goal, studied the human tutor and attempted to absorb and adapt what they learned into Intelligent Computer-Assisted Instruction (ICAI) or Intelligent Tutoring Systems (ITS) \cite {c2}.  

Self, in a paper published in 1990, claimed that ITSs should be viewed as an engineering design field. Therefore, ITS design should be guided by methods and techniques appropriate for design   \cite{c3}\cite{c4}. Twenty years after Self's claim, ITSs had become a growing field with signs of vitality and self-confidence \cite{c3}. 
			
			Intelligent tutoring systems  motivate students to perform challenging reasoning tasks by capitalizing on multimedia capabilities to present information. ITSs have successfully been used in all educational and training markets, including homes, schools, universities, businesses, and governments. One of the goals of  ITSs is to better understand student behaviors through interaction with students \cite {c2}.
			
\begin{figure}[!t]
				\centering
				\includegraphics[width=3in]{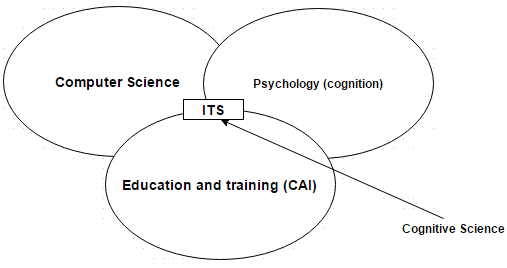}
				\caption{The Domain of ITS Adapted From \cite{c6}.}
				\label{fig1}
			\end{figure}

			ITSs are computer programs that  use AI techniques to provide intelligent tutors that know what they teach, whom they teach, and how to teach. AI helps simulate human tutors in order to produce intelligent tutors. ITSs differ from other educational systems such as Computer-Aided Instruction (CAI). A CAI generally lacks the ability to monitor the learner's solution steps and provide instant help \cite {c211}.  For historical reasons, much of the research in the domain of educational software involving AI has been conducted under the name of Intelligent Computer-Aided Instruction (ICAI). In recent decades, the term ITS has often been used as a replacement for ICAI.  The field of ITS is a combination of computer science, cognitive psychology, and educational research (Figure \ref{fig1}). The fact that ITS researchers use three different disciplines warrants important consideration regarding the major differences in research goals, terminologies, theoretical frameworks, and emphases among ITS researchers. Consequently, ITS researchers are required to have a good understanding of these three disciplines, resulting in competing demands. Fortunately, many researchers have stood up to meet this challenge  this challenge  \cite{c5}\cite{c6}\cite{c7}.

\section{Related Survey Papers}
			The field of ITS has a long history of productive research and continues to grow. There have been a number of well known surveys to keep researchers, new and old, updated. In this section, we will list these surveys with a few key points about each. These surveys can be divided into two main categories. The first category belongs to the surveys that present a general discussion of ITSs. The second category belongs to the surveys that specialize in a specific dimension in ITS.
			 
			A well-known survey which belongs to the first category was published in 1990 by Nwana \cite{c6}. The survey identifies components of ITSs and describes the evolution from Computer-Assisted Instruction and some of popular ITSs of that era. Another survey on ITS was published in 1994 by Shute et al. \cite{c203}. This  is a more in-depth survey regarding the history of ITS, ITS evaluation and the future  of ITSs as seen at the time. Finally,  in-depth case studies were published by Woolf et al. in 2001 \cite {c220} for the purpose of presenting  intelligent capabilities of ITSs when interacting with students. Four tutors were used to exhibit these abilities. The authors ended by discussing evaluations, and some critical development issues for ITSs of the time. 
			
			The other survey category  is more concerned with reviewing a specific dimension of ITSs. Authoring tools in ITSs were reviewed by Murry and Tom in 2003 \cite{c190}. The paper is an in-depth summary and analysis of authoring tools in ITSs along with a characterization of each authoring tool. Another example of  a specific topic-based survey paper is in regard to conversational ITSs \cite{c164}. The history of constraint-based tutors were reviewed in 2012 by Mitrovic \cite{c28}. The paper concentrated on the history and advanced features that have been implemented in tutoring systems. Other survey papers in this category covers dimensions such as behavior of ITSs \cite{c211}, and behavior of ITSs in ill defined domains \cite{c35}.

\section{Why This Survey} 
		    The study of ITSs is a considerable research area as it involves a large number of researchers, working on topics that have strong relations with other disciplines. Interested beginners to this field may struggle to understand the basic aspects and methodologies of ITSs. Difficulties for beginners include understanding how an ITS generally works, AI technologies involved and their functions, learning theories and their uses, main ITS types and how they are different in terms of interaction behaviors, and the importance of ITSs in education and how effective they are.             
			
			The main goal of this work is to provide interested beginners with an updated, in-depth and demystifying introduction to  the field. It is an extensive literature review that presents the main aspects of ITSs in a limited number of pages and direct the readers to the appropriate old and recent references. The paper neither aims to focus on a specific topic or dimension in ITSs, nor does it envision details that may require  hundreds of pages. After understanding ITSs' main concepts and behaviors, a reader can move on to extensively detailed, but slightly dated sources such as Woolf \cite{c212} to take the next steps.

			\section{HUMAN TUTORS VS. COMPUTER TUTORS }
			A number of studies have shown the effectiveness of one-on-one human tutors \cite{c215}\cite{c216}\cite{c141}. When students struggle with difficulties in understanding concepts or exercises, the most effective choice is to seek a one-on-one tutor. There are a variety of features that human tutors are able to provide to students. Good human tutors allow students to do as much work as possible while guiding them to keep them on track towards solutions \cite{c11}. Of course, students learning by themselves also can increase their knowledge and reasoning skills. However, this may consume much  time and effort. A one-on-one tutor allows the student to work around difficulties by guiding them to a strategy that works and helping them understand what does not. In addition, tutors usually promote a sense of challenge, provoke curiosity, and maintain a student's feeling of being in control. Human tutors give hints and suggestions to students rather than giving them explicit solutions. This motivates students to overcome challenges. Furthermore, human tutors are highly interactive in that they give constant feedback to students while the students are solving problems. In order to enable an ITS to give similar feedback as given  by a human tutor, we must ensure that it interacts with students as human tutors do. This leads to the question of how to make an ITS deal with students as effectively as human tutors.

			When modeling ITSs, a student's problem solving processes must be monitored step by step. By keeping track of the steps incrementally, it is possible to detect if a student has made a mistake so that the system can intervene to help the student recover. Feedback can be provided when mistakes are made and hints can be given if students are unsure of how to proceed. One technique used for tracing a student's problem solving is to match the steps a student takes with a rule-based domain expert. In the model tracing technique, the system monitors and follows a student's progress step by step. In case the student makes an error or a wrong assumption, the system intervenes to give explanatory feedback, a hint, or a suggestion to allow the student diagnose  errors. Otherwise, the system silently follows the student's progress. A lot of experiments have shown how student model tracing facilitates learning performance in many educational areas such as the visual presentation in the Geometry Tutor and the graphical instruction in the LISP tutor (GIL) \cite{c11}. Indeed, model tracing tutoring systems support students' learning of the target domain  \cite{c11} \cite{c12}.

	\section{Effectiveness of ITS }
			
			An important question to answer is whether or not  ITSs are really effective in providing the learning outcomes they claim to obtain. There have been a number of meta-analysis efforts to investigate the effectiveness of ITSs. The following present a few recent such efforts with their findings to answer the question. 
			
			A meta-analysis was conducted by VanLehn  in 2011 for the purpose of comparing effectiveness of computer tutoring, human tutoring and no tutoring \cite{c12}. In this analysis,  computer tutors were characterized based on the granularity of the user interface interactions, including answer-based, step-based, and substep-based tutoring systems. Their analysis included studies published between 1975 and 2010. 10 comparisons were presented from 28 evaluation studies. The study found that human tutoring raised  test scores by an effect size of 0.79 compared to no tutoring; thus it is not as effective as 2.0 found by Bloom earlier \cite{c141}. Moreover, it was found that step-based tutoring (0.76) was as almost effective as human tutoring whereas substep-based tutoring was only 0.40 as effective compared to no tutoring. VanLehn's findings suggest that tutoring researchers should focus on ways to improve computer tutoring to reach up to  Bloom's finding that human tutoring has 2.0 multiplicative effect compared to no tutoring.

			The meta-analysis conducted by Steenbergen and Cooper in 2013 analyzed the effectiveness of ITSs on k-12 students' math learning \cite{c208}. This empirical research examined 26 reports comparing the effectiveness of ITSs with that of regular classroom instruction. Their finding was that ITSs did not have a significant effect on student learning outcomes when used for a short period. However, the effectiveness appeared to be greater when ITS was used for one full school year or longer. In addition, the effects appeared to be greater on general students than on low achievers.

			The meta-analysis by Ma et al. \cite {c209} was conducted in 2014 for the purpose of comparing the learning outcomes for those who learn by using ITSs and those who learn in non-ITS learning environments. Their goal was to verify the effect sizes of ITSs taking into account factors such as type of ITS, type of instruction (individual, small, large human instruction etc.), and subject domain (chemistry, physics, mathematics etc.), and other factors. Ma et al., analyzed 107 effect size findings from 73 separate studies. The ITS environment was associated with greater learning achievement compared to teacher-led and large group instruction with an effect size of 0.42, 0.57 for non-ITS computer based instruction, and 0.35 for text books or workbooks. On the other hand, there was no considerable difference between learning outcomes from ITSs and from individualized human tutoring (-0.11) or small group instruction (0.05). Ma et al., reported that ITSs achieved higher education outcome than other forms of instructions except for small group human tutoring. In addition, the ITS effect varied as features and characteristics of ITSs, student attributes, domain knowledge, and other factors varied.
			
			Finally, the meta-analysis produced by Kulik and Fletcher in 2015 \cite {210} compared the learning effectiveness of ITSs with conventional classes from 50 studies. 92\% of the  studies indicated that students who interacted with ITSs outperformed  those who received traditional class instructions. In 39 of the 50 studies, performance improvement gains were up to 0.66 median effect sizes, which is considered to be moderate to strong. However, the effect was weak for standardized tests as the effect size was 0.13. 
			
			Because of the fact that there is no general agreement on the effectiveness of ITSs, questions come up for  researchers to answer. How effective are ITSs really?, What are the critical reasons that affect  learning in ITSs?, What possible changes can be made to improve ITSs?

			\section{ARCHITECTURE OF ITS }
\begin{figure}[!t]
				\centering
				\includegraphics[width=3in]{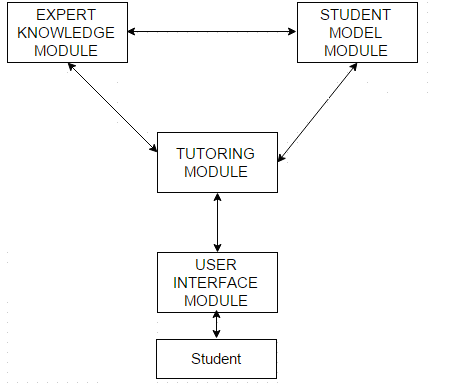}
				\caption{Traditional Architecture of ITS (Adapted from \cite{c6}).}
				\label{fig2}
			\end{figure}

			 ITSs vary greatly in architecture. It is very rare to find two ITSs based on the same architecture. There are three types of knowledge that ITSs possess: knowledge about the content that will be taught, knowledge about the student, and knowledge about teaching strategies. Additionally, an ITS needs to have communication knowledge in order to present the desired information to the students. Consequently, the traditional `typical' ITS has four basic components: the domain model which stores domain knowledge, the student model which stores the current state of an individual student in order to choose a suitable new problem for the student, and the tutor model which stores pedagogical knowledge and makes decisions about when and how to intervene. The intervention can use different forms of interactions: Socratic dialogs, hints, feedback from the systems, etc. Finally the user interface model gives access to the domain knowledge elements. Figure \ref{fig2} shows the traditional architecture of ITSs \cite{c3}\cite{c6}\cite{c13}.
			
			In addition, even though ITSs differ greatly in their internal structures and components and contain a wide variety of features, their behaviors are similar in some ways as stated by VanLehn \cite{c211}. According to VanLehn, ITSs behave in similar ways such that they involve two loops named inner loop and outer loop. The outer loop mainly decides which task students should practice next among other tasks. The decision takes place based on the student's history of knowledge and background. The inner loop is responsible for monitoring the student's solution steps within a task by providing appropriate pedagogical intervention such as feedback on a step, hints on the next step, assessment of knowledge and review of the solution.

			The goal of this section is to describe these components along with their functions.
			
\subsection{Domain Model }
			The expert knowledge, the domain expert, or the expert model,  represents the facts, concepts, problem solving strategies, and rules of the particular domain to be taught, and provides ITSs with the knowledge of what they are teaching. The material and detailed knowledge are usually derived from experts who have years of experience in the domain to be taught. It is important to mention that to find what to teach is the goal of the domain model. However, it is separate from the control information (how to teach), which is represented by the tutoring model \cite{c18}. The domain expert fulfills a double function. Firstly, it acts as the source of the knowledge to be presented to students through explanations, responses and questions. Secondly, it evaluates the student's performance. In order to accomplish these tasks, the system must be able to present correct solutions to problems so that the student's answers can be compared to those of the system. In case the ITS is required to guide the student in solving problems, the expert model must be able to generate sensible and multiple paths of solutions to help fill the gap in the student's knowledge. The expert model can also provide an overall progress assessment of students by establishing specific criteria with which to compare knowledge \cite{c6}\cite{c15}\cite{c16}\cite{c17}.
			
			An ITS must have a knowledge base system which contains information on what will be taught to the learners. The need for suitable knowledge representation (KR) languages must be considered in representing and using the knowledge. The principles that need to be considered when choosing KR languages to build the knowledge are the expressivity of the language, the inference capacity of the language, the cognitive plausibility of the language and pedagogical orientation of the language \cite{c3}. Hatzilygeroudis and Prentzas made the first efforts to define and analyze the requirements for knowledge representation in ITSs \cite{c14}. Various knowledge representation and reasoning schemes have been used in ITSs. These include symbolic rules, fuzzy logic, Bayesian networks, and case-based reasoning, and hybrid representations such as  neuro-symbolic and neuro-fuzzy approaches. More details on examples of ITS systems along with the knowledge representation languages used can be found in \cite{c3}. 
			
			The following explains three traditional types of ITS approaches for representing and reasoning with the domain knowledge. Two types of domain knowledge models used frequently in ITSs are the cognitive model, and the constraint based model. The third approach incorporates an expert system in the ITS \cite{c28}.
			
			\subsubsection{Cognitive Model}
\begin{figure}[!t]
				\centering
				\includegraphics[width=3in]{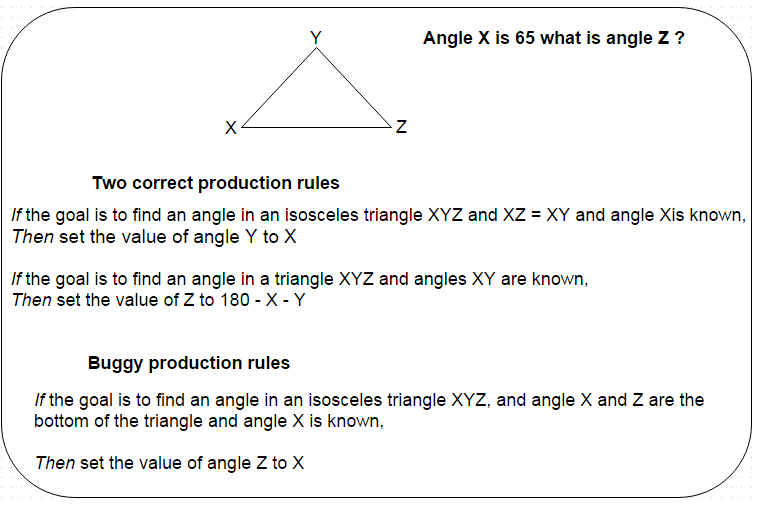}
				\caption{Production and Buggy Rules for Computing the Size of an Angle (Adapted From \cite{c23}).}
				\label{fig3}
			\end{figure}

			The cognitive model is a traditional approach to model the domain knowledge in  ITSs. It has been used in a family of successful ITS systems \cite{c19}. The tutors that use a cognitive model in designing the tasks in the domain have been called Cognitive Tutors. Cognitive tutors are effective and several scientific studies have found that cognitive tutors improve student learning and demonstrated learning outcomes \cite{c20}. They have been fielded in a variety of scientific domains such as algebra, physics, geometry and computer programming \cite{c21}. Cognitive tutors use a cognitive model to provide students with immediate feedback. The goal of this approach is to provide a detailed and precise description on the relevant knowledge in a task domain including principles and strategies for problem-solving. A rule-based model generates a step by step solution to provide support to students in a rich environment for problem solving that generates feedback to students on the correctness of each step in the solution and can keep track of many approaches (strategies) to the final correct answers. Not only are the correct solutions  represented, but also the common mistakes that the students usually make (Bug Libraries) as shown in Figure \ref{fig3}  \cite{c22}\cite{c23}.
			
			Cognitive tutors have been built based on the ACT-R theory of cognition and learning \cite{c3}. The underlying principle of ACT-R is the distinction between explicit and implicit knowledge. Procedural knowledge is considered implicit whereas declarative knowledge is explicit. Declarative knowledge consists of facts and concepts in the form of a semantic net or similar network of concepts linking what are called chunks. In contrast, procedural memory represents knowledge of how we do things in the form of production rules written in \textit{IF-THEN} format. Thus, chunks and productions are the basic forms of an ACT-R model \cite{c2}\cite{c63}.

			In order to use cognitive models to facilitate tutoring, an algorithm called model tracing has been used. The tutor assesses the student solution by comparing the student solution steps against what the model would do for the same task. If the student action is the same as the model action, it is deemed correct. Otherwise it is not correct. An error is hypothesized when a student step does not match any rule or it matches one or more of the buggy rules \cite{c5}. Each production rule that generates the matching action can be interpreted as a skill possessed by the student. So over time, the model is able to evaluate the skills that have been mastered by the student (knowledge tracing). Thus, knowledge tracing is used to monitor the skills that students have acquired from solving a problem \cite{c63}\cite{c25}.
			
			The Knowledge Tracing model  called Cognitive Mastery Learning is one of the most popular methods for estimating the probability that a student knows each skill \cite{c217}. The model continuously keeps  assessing the probability that a student has acquired each skill taking into account four parameters for each skill. Cognitive Mastery Learning is known to produce a significant improvement in learning and it has a long history of application. Educational data mining approaches such as Learning Factor Analysis (LFA) \cite{c218} and Performance Factors Analysis (PFA) \cite{c219} have been used to futher improve  ITSs using this model.          
			
			Despite the fact that cognitive tutors have led to impressive student learning gains in a variety of domains, these model tracing tutors have not been widely adopted in educational or other settings such as corporate training. The fact is that building complete and optimal cognitive tutors requires software pieces such as an interface, a curriculum, a learner interacting management system, and a teacher reporting package. Additionally, the process also needs a team of professionals to work together, resulting in high cost and time. These two requirements have limited practical use of such tutors \cite{c26}. To reduce the cost of building model tracing tutors, authoring tools with some of the capabilities have been built. An example is  Cognitive Tutor Authoring Tools (CTAT) \cite{c27}. CTAT has been used to create  full ITSs without programming. This has led to a new paradigm of an ITS called Example-Tracing Tutors \cite {c27}.

			\subsubsection{Constraint based Model (CBM)}
			Constraint based modeling was proposed by Ohlson in 1992 to overcome difficulties in building the student model  \cite{c28}. Since then, CBM has been used widely in numerous ITSs to represent instructional domains, students' knowledge and higher level skills. CBM is based on Ohlson's theory of learning using performance errors, resulting in a new methodology for representing the knowledge using constraints which cannot be violated during problem solving. It is different from model tracing, which generates all possible paths of solutions using production rules. CBM can be used to represent both domain and student knowledge. This model has been used to design and implement efficient and effective learning environments \cite{c28}\cite{c29}.
			
			The fundamental idea behind CBM is that constraints represent the basic principles and facts in the underlying  domain which a correct solution must follow \cite{c30}. The observation here is that all correct solutions for any problem are similar in that they do not violate any domain principles or ``constraints". Instead of representing both correct and incorrect spaces as in model tracing, it is sufficient to capture only the domain principles \cite{c28}.

The form of a constraint is an ordered pair (Cr, Cs), where Cr is the relevance condition and Cs is the satisfaction condition so the constraint follows the form:
\begin{quote}
			If \textit{$<relevance$  $condition>$} is true,\\
			Then \textit{$<satisfaction $ $condition>$} had better also be true.
			\end{quote}
			
			The relevance condition may contain simple or compound tests to specify  features of student solutions whereas the satisfaction condition is an additional test that has to be met in order for the student's solution to be correct \cite{c21}. 
			
			The CBM approach was proposed to avoid some limitations of model tracing tutors. First, the nature of CBM's knowledge representation as constraints allows for creativity. The system accepts student's solutions even though it is not represented in the system as long as they do not violate any constraints.  On the contrary, model tracing limits the students' solutions to ones stored in the model. Thus, the idea that different students might have different strategies or beliefs to find their results is considered. Second, creating a bug library as used in model tracing requires a lot of time since the types of mistakes can be vast. Consequently, CBM concentrates on domain principles that every correct solution must meet. CBM's hypothesis in this regard is that all correct solutions share the same features, so it is enough to represent the correct space by capturing domain principles. In case of student errors, the system can advise the student on the mistake without being able to represent it. Finally, for some instructional tasks categorized as ill-defined (for details see \cite{c35}), it may even be impossible to follow the steps of the correct solutions because the runnable models are expressed as set of production rules for both the expert and student. CBM avoids this limitation and can handle some ill-defined tasks \cite{c29}.

			\subsubsection{Expert Approach}
			The third approach for representing and reasoning with domain knowledge consists of integrating an expert system in an ITS. This is considered a broad approach in ITS since several formalisms of expert systems can be used such as rule-based, neural networks, decision trees and case-based reasoning. An expert system mimics the ability of an expert in terms of decision making and skills in modeling, and solves a problem \cite{c36}.  The advantage of the expert system approach is in its ability to accept a broader domain to represent and reason with, unlike constraint based and cognitive models work with limited  domains \cite{c35}.
			
			Fournier-Viger et al. \cite{c35} showed that an expert system approach should provide for two modalities. First, the expert system should be able to generate expert solutions and then compare these solutions with the learner's solutions. GUIDON \cite{c137} is as an example of an ITS that uses this modality. The second modality for using an expert system approach in ITSs is to compare ideal solutions with the learners' solutions. Some examples of systems that are able to meet the second modality are AutoTutor \cite{c138}, and DesignFirst-ITS \cite{c139}.
			
			Despite the fact that the expert system approach is powerful, it faces some limitations as noted by \cite{c35}: ``(1) developing or adapting an expert system can be costly and difficult, especially for ill-defined domains; and (2) some expert systems cannot justify their inferences, or provide explanations that are appropriate for learning''.  
			
\subsection{Student Model}
			
			It would be difficult for an ITS to succeed without some understanding of the user. The student model represents the knowledge and skills of the student dynamically. Just as domain knowledge must be explicitly represented so that it can be communicated, the student  model must also be represented likewise. Ideally, the student model should  store aspects of the student's behavior and skills in such a way that the ITS can infer the student's performance and skills.
			
			According to Nwana, the uses of the student model can be classified into six different types \cite{c6}. The first type is corrective in that it enables removing bugs in a student's knowledge. The second type is elaborative in that it fills in the student's incomplete knowledge. The third type is strategic in that it assists in adapting the tutorial strategy based on the student's action and performance. The fourth type is diagnostic in that it assists in identifying errors in the student's knowledge. The fifth type is predictive in that it assists in understanding the response of the student to the system's actions. The sixth and final type is evaluative in that it assists in evaluating the students overall progress.

			The student model acts as a source of information about the student. The system should be able to infer unobservable aspects of the student's behavior from the model. It should reconstruct  misconceptions in the student's knowledge by interpreting the student's actions. The representation of the student model is likely to be based on the representation of domain knowledge. The knowledge can be separated into elements with evaluations of mastery incorporated into the student model. This allows the system to compare the state of the student's knowledge with that of the expert. As a result, instructions can be adapted to exercise the weaknesses in the student's skills. It should be noted that incomplete knowledge is not necessarily the source of incorrect behavior. The knowledge to be taught can evolve, which presents a challenge to the tutoring system. It is for this reason that explicit representations of a student's supposed incorrect knowledge must be included in a student model so that remediation can be performed as necessary. An important feature of the student model is that it is executable or runnable. This allows for prediction of a particular student's behavior in a particular context. This ultimately allows this important architectural component of an ITS to interact appropriately with the student. These interactions may include correction of misconceptions, providing personalized feedback, and suggestion for learning a particular item, etc. \cite{c6} \cite{c7}.
			
			Designing a student model is not an easy mission. It should be based on responses to certain questions. What does the student know? What types of knowledge will the student need to solve a problem? It is from such questions that the methodology for designing a student model should derived. It is first necessary to identify the knowledge that the student has gained in terms of the components that are integrated with the mechanism. It is secondly necessary to identify the understanding level of the student vis-a-vis the functionality of the mechanism. It is finally necessary to identify the pedagogical strategies used by the student to arrive at a problem's solution. These  must be taken under consideration in the development of the student model \cite{c16}. 
			
			There are several kinds of student characteristics that should be taken into consideration. In order to build an efficient student model, the system needs to consider both static and dynamic characteristics of students. Static characteristics include information such as email, age, and mother tongue, and are set before the learning processes start, whereas dynamic characteristics come from the behavior of students during the interaction with the system \cite{c39}\cite{c40}. According to \cite {c39}, the challenge is to find the relevant dynamic characteristics of an individual student in order to adapt the system for each student. The dynamic characteristics include knowledge and skills, errors and misconceptions, learning styles and preferences, affective and cognitive factors, and meta-cognitive factors. The term knowledge here refers to the knowledge that has been acquired by the student previously, while learning style or preferences refer to how the student prefers to perceive the learning material (e.g., graphical representation, audio materials and text representation). Affective factors include the emotional characteristics of the students such as being angry, happy, sad, or frustrated. Cognitive factors refer to the cognitive features of students, for instance, attention, ability to learn and ability to solve problems and make decisions. Meta-cognitive aspects involve attitude and ability for help-seeking, self-regulation, and self-assessment \cite{c39}\cite{c41}.

			Several approaches  have been used to build the student model. The following subsections discuss some approaches that have been found in the literature.
			
			\subsubsection{Overlay Model}
			The overlay model was invented in 1976 by Stansfield, Carr and Goldstein. It is one of the most popular student models, and it has been used by many tutoring systems. It assumes that student knowledge is a subset of domain knowledge. If the student has a different behavior from that of the domain, it is considered a gap in the student's knowledge. As a result, the goal is to eliminate the gap between them as much as possible \cite{c39}\cite{c42}. Consequently, the domain contains a set of elements and the overlay model indicates a set of masteries over these elements. A simple overlay model uses  a Boolean value to indicate if an individual student knows the element or does not know the element. In the modern overlay model, a qualitative measure is used to indicate the level of student knowledge (good, average or poor). The advantage of using this model is that it allows making the amount of student knowledge as large as necessary. However, the disadvantage of using this model is that the student may take a different approach to solve a problem. The student may also have different beliefs in the form of `misconceptions' that are not stored in the domain knowledge \cite {c39}. 
			
			Carmon and Conejo in 2004 proposed a learner model in their MEDEA system, which is a framework to build open ITSs \cite{c44}. The classical overlay model was used to represent knowledge and attitude of the students in MEDEA. The learner model was divided into two sub-models: attitude model and learner knowledge model. The attitude model contains static information about the students (users' personal and technical characteristics, users' preferences, etc.). These features were collected directly from the student before the learning processes take place. The learner knowledge model was responsible for the student's knowledge and performance. These features were updated during the learning processes. For each domain concept, the learner model stores an estimation of the knowledge level of the student on this concept \cite{c42}\cite{c44}.
			
			InfoMap is designed to facilitate both human browsing and computer processing of the domain ontology in a system. It uses the overlay model combined with a buggy model to identify deficient knowledge \cite{c45}. Another ITS that applies the overlay model for the student model is ICICLE (Interactive Computer Identification and Correction of Language Errors). The system's goal is to employ natural language processing to tutor students on grammatical components of  written English. ICICLE uses a student overlay model to capture the user's mastery of various grammatical units and thus can be used to predict the grammar rules he or she is most likely using when producing language \cite{c46}. Kumar in 2006 used the overlay student model in an ITS for computer programming called DeLC (Distributed eLearning Center) for distance and electronic teaching \cite{c47}. It used the overlay student model to capture the level of understanding of the user. However, it also used another modeling approach named the stereotype approach to model learner's manner of access to training resources, their preferences, habits and behaviors during the learning process \cite{c48}. 
			
			LS-Plan is a framework for personalization and adaption in e-learning systems. It uses a qualitative overlay model based on Bloom's Taxonomy. LS-Plan also uses a bug model to detect misconceptions of the users \cite{c49}. PDinamet, a web-based adaptive learning system for the teaching of physics in secondary school, uses an overlay student model to store concepts that the student has already learned or has not learned yet. Consequently, the tutor can recommend to an individual student a certain topic by taking into account the student's skill level and the learning activities the student has already participated in PDinamet \cite{c50}.\\
			
			The overlay model neither takes into account the incorrect knowledge that the student has nor the students' cognitive needs, preferences and learning styles. This is the reason why most personalized tutoring systems combine the overlay model with other approaches such as stereotypes, fuzzy logic, machine learning, and perturbation \cite{c39}.
			
			\subsubsection{Stereotypes Model}
			Another widely used approach for student modeling is in terms of stereotypes. The stereotypes approach in student modeling began with a system called GRUNDY by Rich \cite{c52}. According to Rich, ``A stereotype represents a collection of attributes that often co-occur in people. They enable the system to make a large number of plausible inferences on the basis of a substantially smaller number of observations. These inferences must, however, be treated as defaults, which can be overridden by specific observation''\cite{c51}\cite{c52}.
			
			The main assumption in stereotypes is that it is possible to group all possible users based upon certain features they typically share. Such groups are called stereotypes. A new user will be assigned to a specific stereotype if his/her features match this stereotype. Most ITSs give students freedom to choose, meaning that the student chooses his/her own learning path in the courseware. As a consequence, students may study material that is too hard or too easy for them, or skip learning certain courseware elements. Beside generating, selecting and sequencing material for the students, the ITSs should take into consideration the current knowledge of the students. As they reduce the cognitive overload as well as provide individualized guidance for learning and the teaching process \cite{c53}, stereotypes are particularly important to overcome the problem of initializing a student model by assigning the student to a certain group of students. The system might ask the user some questions to initiate its student model \cite{c39}. 
			
			For example, let us consider a system that teaches the Python programming language. The system might start interactions with students by asking questions in order to discover the stereotype this student belongs to. A related question that could be asked is if the student is an expert in C++ programming. If the student is an expert in C++,  the system would infer that this student knows the basic concepts in programming such as loops, while loops and nested loops. Consequently, the system will assign this particular student to a stereotype whose members know these basic programming concepts \cite{c51}.

			Many adaptive tutoring systems have used the stereotype approach to student modeling and often combine them with other student modeling approaches \cite{c51}. INSPIRE is an ITS for personalized instruction. The stereotype approach is used to classify knowledge on a topic to one of four levels of proficiency: Insufficient, Rather Insufficient, Rather Sufficient, Sufficient.  Besides stereotypes, the fuzzy logic technique is used to deal with student diagnosis \cite{c54}. Another ITS using stereotypes is Web-PVT which teaches the passive voice in English using the Web Passive Voice Tutor. Machine learning and stereotypes were used to tailor instruction and feedback to each individual student. The initialization of the model for a new student is performed using a novel combination of stereotypes and a distance weighted k-nearest neighbor algorithm \cite{c55}.
			
			AUTO-COLLEAGUE, an adaptive computer supported collaborative learning system, aims to provide a personalized and adaptive environment for users to learn UML  \cite{c57}. AUTO-COLLEAGUE uses a hybrid student model that combines stereotypes and the perturbation approach, to be discussed next. The stereotypes are concerned with three aspects of the user (the level of expertise, the performance type and the personality). Another ITS that uses stereotype for student modeling is CLT. CLT teaches C++ iterative constructs (while, do-while, and for loops). The triggers for the stereotypes used in CLT are verbal ability, numerical ability, and spatial ability, each of which can be rated low, medium and high \cite{c58}.
			
			According to Chrysafiadi et al., the advantages of the stereotypes technique are that the knowledge about a particular user is inferred from the related stereotype(s) as much as possible, without explicitly going through the knowledge elicitation process with each individual user \cite{c39}. In addition, maintaining information about stereotypes can be done efficiently with low redundancy. On the other hand, the disadvantages of stereotypes are that stereotypes are constructed based upon external characteristics of users and on subjective human judgment, usually of a number of users/experts. It is common that that some stereotypes do not represent their members accurately. Therefore, many researchers have pointed out the issue of inaccuracy in stereotypes. Stereotypes suffer from two additional problems. First, the users must be divided into classes before the interactions with the system begin, and as a result, some classes might not exist. Second, even if a class exists, the designer must build the stereotype, which is time consuming and error-prone.

			\subsubsection{Perturbation Model}
			The perturbation student model is an extension of the overlay model. Besides representing the knowledge of the students as a subset of the expert's knowledge (overlay model), it also includes possible misconceptions which allow for better remediation of student mistakes \cite{c59}. The perturbation model incorporates misconceptions or lack of knowledge, which may be considered mal-knowledge or incorrect beliefs \cite{c60}.  According to Martins et al., the perturbation model can be obtained by replacing correct rules with wrong rules \cite{c59}. When applied, they produce the answers given by the student. Since there can be several reasons for a student's wrong answer (several wrong rules in student knowledge before the beginning of the interaction with the student to acquire knowledge and after interaction with specialized knowledge),  the system proceeds to generate discriminating problems and presents them to the student to identify the wrong rules that this user has.
			
			The mistakes that students make are usually stored in what is termed the bug library. The bug library is built by either collecting the mistakes that students make during interaction with the system (enumeration) or by listing the common misconceptions that students usually have (generative technique). This model gives better explanations of student behavior than the overlay model. However, it is costly to build and maintain \cite{c40}.

			Many adaptive tutoring systems have used the perturbation technique for their student model. Hung and his colleagues in 2005, used the perturbation model (also called buggy model), with 31 types of addition errors and 51 subtraction errors, to help the system analyze and reason with student's mistakes \cite{c45}. LeCo-EAD uses the perturbation model to represent students' incorrect knowledge to provide personalized feedback and support to distant students in real time \cite{c61}. The perturbation model was also used by Surjano and Maltby in combination with stereotypes and the overlay model to perform a better remediation of student mistakes \cite{c62}. Baschera and Gross also used a perturbation student model in 2010 for the purpose of spelling training to better diagnose students' errors \cite{c36}.
			
			\subsubsection{Constraint Based Model}
			
			The constraint based model (CBM) was first published for short-term student modeling and the diagnosis of the current solution state. CBM uses constraints to present both domain and student knowledge \cite{c39}. The process of diagnosing the student's solution is by matching the relevance conditions of all constraints to the students' solutions. The satisfaction condition for all relevance conditions are matched as well. The system checks each step taken by the student, diagnoses any problem, and provides feedback to the student when there is an error. The feedback informs the student that the solution is wrong, indicates the part of the solution that's wrong, and then specifies the domain principle that is violated \cite{c28}. According to Mitrovic et al., important advantages of CBM are that CBM does not require a runnable expert module, leading to computational simplicity; it does not require extensive studies of students' bugs; and it does not require complex reasoning about possible origins of student errors \cite{c64}. These advantages have led researchers to apply the CBM approach to their tutoring systems in a variety of domains.

			SQLT-Web is a web enabled ITS for the SQL database language.  It diagnoses the student's solution and adapts feedback to his/her knowledge and learning abilities \cite{c65}.  J-LATTE, which is an ITS that teaches a subset of the Java programming language, uses the CBM approach in the student model. When the student submits his/her solution, the student modeler evaluates it and produces a list of relevant, satisfied and (possibly) violated constraints \cite{c100} \cite{c66}. INCOM is a system which helps students of a logic programming course at the University of Hamburg. Weighted constraints are used to achieve accuracy in diagnosing students' solutions \cite{c67}. EER-Tutor is also another system that teaches database concepts and adapts the CBM student model to represent the student's level of knowledge \cite{c28}.
			
			\subsubsection{Cognitive Theories}
			The use of cognitive theories for the purpose of student modeling and error diagnosis leads to effective tutoring systems, as many researchers have pointed out. A cognitive theory helps interpret human behavior during the learning process by trying to understand human processes of thinking and understanding. The Human Plausible Reasoning (HPR) Theory \cite{c68}, and the Multiple Attribute Decision Making (MADM) Theory \cite{c69} are some cognitive theories that have been used in student modeling \cite{c39}.

			Human Plausible Reasoning (HPR) is a theory which categorizes plausible human inferences in terms of a set of frequently recurring inference patterns and a set of transformations on these patterns. In particular, it is a domain-independent theory, originally based on a corpus of people's answers to everyday questions \cite{c68}. A system that uses HPR in student modeling is RESCUER, which is an intelligent help system for UNIX users. The set of HPR transformations are applied to statements to generate different possible interpretations of how a user may have come to the conclusion that the command s/he typed is acceptable to UNIX \cite{c70}. Another system that uses HPR to model the student is F-SMILE. F-SMILE stands for File-Store Manipulation Intelligent Learning Environment, which aims to teach novice learners how to use file-store manipulation programs. The student model in F-SMILE captures the cognitive state, as well as the characteristics of the learner and identifies possible misconceptions. The LM Agent in F-SMILE uses a novel combination of HPR with a stereotype based mechanism to generate default assumptions about learners until it is able to acquire sufficient information about each individual learner \cite{c71}.

			Another cognitive theory which has been used to build student models is Multiple Attribute Decision Making (MADM) \cite{c69}. MADM makes preference decisions (e.g., evaluation, prioritization, or selection) among available alternatives that are usually characterized by multiple, usually conflicting, attributes.  Web-IT is a Web-based intelligent learning environment for novice adult users of a Graphical User Interface (GUI) that manipulates files, such as the Windows 98/NT Explorer. Web-IT uses MADM in combination with an age stereotype to dynamically provide personalized tutoring \cite{c72}. A novel mobile educational system has been developed by Alepis and Kabassi, incorporating bi-modal emotion recognition based on two modes of interactions,  mobile device microphone and  keyboard, through a multi-criteria decision making theory to improve the system's accuracy in recognizing emotions \cite{c73}.

			\subsubsection{Bayesian Networks}
			
			Another well-known and established approach for representing and reasoning about uncertainty in student models is Bayesian networks \cite{c39}. A Bayesian network (BN) is a directed acyclic graph containing random variables, which are represented as nodes in the network. A set of probabilistic relationships between the variables is presented as arcs. The BN reasons about the situation it models, analyzing action sequences, observations, consequences and expected utilities \cite{c40}. Regarding the student model, components of students such as knowledge, misconceptions, emotions, learning styles, motivation and goals can be represented as nodes in a BN \cite{c39}. 
			
			BNs have been shown to be powerful and multi-purpose when modeling any problems that involve knowledge. Bayesian networks have attracted attention from theoreticians and system designers not only because of sound mathematical foundation, but also for being a natural way to represent uncertainty using probabilities. Therefore, BNs have been used in many different domains such as medical diagnosis, information retrieval, bioinformatics, and marketing,  for many different purposes such as troubleshooting, diagnosis, prediction, and classification \cite{c40}. Those who are interested in using Bayesian networks can use tools such as GeNIe \cite{c143} and SMILE \cite{c144} for easy creation and efficient BNs.
			
			Andes is an ITS providing help in the domain of Newtonian physics \cite{c75}\cite{c76}. The student model in Andes uses Bayesian networks to carry out long-term knowledge assessment, plan recognition, and prediction of the students' actions during problem solving. Another student model that uses BN is in  Adaptive Coach Exploration (ACE), an intelligent exploratory learning environment for the domain of mathematical functions. The student model is capable of providing tailored feedback on a learner's exploration process, also detecting when the learner is having difficulty exploring \cite{c77}. A Bayesian student model also has been implemented in the context of an Assessment-Based Learning Environment for English grammar. A Bayesian student model is used by pedagogical agents to provide adaptive feedback and adaptive sequencing of tasks \cite{c78}.
			
			A Bayesian student model is also used in E-teacher to provide personalized assistance to e-learning students with the goal of automatically detecting a student's learning style \cite{c79}. A Dynamic Bayesian network was used by Conati and Maclaren to recognize a high level of uncertainty regarding multiple user emotions by combining information on both the causes and effects of emotional behavior \cite{c80}. Similarly, a Dynamic Bayesian network was implemented in PlayPhysics to reason about the learner's emotional state from cognitive and motivational variables using observable behavior \cite {c81}. TELEOS (Technology Enhanced Learning Environment for Orthopedic Surgery) used a Bayesian student model to diagnose the students' knowledge states and cognitive behaviors \cite{c82}. A Bayesian student model was also applied in Crystal Island, which is a game-based learning environment in the domain of microbiology to predict student affects by modeling students' emotions \cite {c83}.
			
			\subsubsection{ Fuzzy student modeling}
			
			In general,  learning and determining the student's state of knowledge are not straightforward tasks, since they are mostly affected by factors which cannot be directly observed and measured, especially in ITSs where there is a lack of real life interaction between a teacher and students. One possible approach to deal with uncertainty is fuzzy logic, introduced by Zadeh in 1956 as a methodology for computing and reasoning with subjective words instead of numbers \cite{c39}. Fuzzy logic is used to deal with uncertainly in real world problems caused by imprecise and incomplete data as well as human subjectivity \cite{c84}. Fuzzy logic uses fuzzy sets that involve variables with uncertain values. A fuzzy set is described by variables and values such as ``excellent'', ``good'' and ``bad'' rather than a Boolean value ``yes/no" or ``true/false''.

			A fuzzy set is determined by a membership function expressed as $U (x)$ \cite{c85}. The value of the membership function $U(x)$ is called the degree of membership or membership value, and has a value between 0 and 1. 
			The use of fuzzy logic can improve the learning environment by allowing intelligent decisions about the learning content to be delivered to the learner as well as tailored feedback that should be given to each individual learner \cite{c39}. Fuzzy logic can also diagnose the level of knowledge of the learner for a concept, and predict the level of knowledge for other concepts that are related to that concept \cite {c84}. Chrysafiadi and Virvou, in 2012, perform an empirical evaluation of the use of fuzzy logic in student modeling in a web-based educational environment for teaching computer programming. The result of the evaluation showed that the integration of fuzzy logic into the student model increases the learner's satisfaction and performance, improves the system's adaptivity and helps the system make more reliable decisions \cite{c86}. The use of fuzzy logic in student modeling is becoming popular since it overcomes computational complexity issues and mimics human-like nature \cite{c39}\cite{c85}.

			\subsection{Tutor Model}
			
			An ITS provides personalized feedback to an individual student based upon the traits that are stored in the student model. The tutor model or the pedagogical module, as it is alternatively called, is the driving engine for the whole system \cite{c87}. This model performs several tasks in order to behave like a human-like tutor that can decide how to teach and what to teach next. The role of the tutor model  is not only to provide guidance like a tutor but also to make the interaction of the ITS with the learner smooth and natural \cite{c88}. The pedagogical module should be able to answer questions such as should the student be presented a concept, a lesson, or a test next. Other questions include deciding how to present the teaching material to the student, evaluate student performance, and provide feedback to the student \cite{c30}\cite{c88}. 
			
			Indeed, the pedagogical module communicates with all other components in the system's expert model, and the student model and acts as an intermediary between them \cite{c89}. When a student makes a mistake, the pedagogical module is responsible for providing feedback to explain the type of error, re-explain the usage of that rule and provide help whenever the student needs it \cite{c90}. The tutor must also decide what to present next to the student such as topic, or the problem to work on. To do so, the pedagogical model must consult the student model to determine the topics on which the student needs to focus. These decisions that this model makes are according to the information about the student stored in the student model and the information about the learned content which the expert model stores \cite{c67}.

			The pedagogical module is responsible for the interaction between the student and the system in case the student needs help at any given step, for remediation of  student's error. It does so by giving a sequence of feedback messages (e.g., hints), or suggesting to the student to study a certain topic to increase learning performance. All ITSs have embedded the pedagogical module to control interaction with the students. The following will present some pedagogical techniques which have been used for the purpose of delivering content and making interventions when needed in ITSs. 
			
			\subsubsection{Decision Making in Cognitive Tutor and Constraint Based Systems}
			
			Model Tracing Tutors (MTT) (Cognitive Tutors), specifically their 2nd generation architecture \cite{c91}, give three types of feedback to students: flag feedback, buggy messages, and a chain of hints. Flag feedback informs the student on the correct or wrong answer by using a color (e.g., green = correct or red = wrong). A buggy message is attached to a specific incorrect answer the student has provided to inform the student of the type of errors s/he has made. 
			
			In case the student needs help, s/he can ask for a hint to receive the first hint from a chain of hints, which include suggestions to make the student think.  The student can request  hints to get more specific hints on what to do, and when the chain of hints is all delivered, eventually, the system tells the student exactly what to type \cite{c22}. CBM tutors such as KERMIT and SQL-Tutor, ITSs that teach conceptual database design, provide six levels of feedback to the student: correct, error flag, hint, detailed hint, all errors and solution. The correct level simply indicates the student whether the answer is correct or incorrect. The error flag indicates the type of construct (e.g., entity and relationship) that contains the error. Hint and detailed hint are feedback that are generated from the violated constraint. The complete solution is displayed at the solution level \cite{c87}\cite {c92}.
			
			Besides providing feedback to remedy students' errors, personalized guidance can also be given to help students, as Kenny and Pahl have done in SQL-Tutor \cite{c93}. They offer the student advice and recommendation about subject areas that a particular student needs to focus on. The decision about the particular areas recommended by the system is determined by collecting data from the student model. The pedagogical model retrieves the information on all errors made by the student from the student model to make the decision.
			
			Model Tracing Tutors have developed teaching strategies and interactions between the system and the student to reach the level of performance of experienced human tutors. However, many researches have criticized model tracing tutor because an MTT needs a strategic tutor \cite{c91}. According to them, an MTT should encourage students to construct their own knowledge instead of telling it to them. In other words, students can learn better if they are engaged in a dialog that helps them construct their knowledge themselves instead of being hinted toward inducing the knowledge from problem-solving experiences. 
			
			A 3rd generation model tracing tutor, named Ms. Lindquist, using what is called an Adding Tutorial Model, was the first model tracing tutor that was designed to be more human-like in caring when participating in a conversation. Ms. Lindquist could produce probing questions, positive and negative feedback, follow-up questions in embedded subdialogs, and requests for explanation as to why something is correct \cite{c91}\cite{c94}.

			DEPTHS, which is an ITS for learning software design patterns, implements a curriculum planning model for selecting appropriate learning materials (e.g., concepts, content units, fragments and test questions) that best fit the  student's characteristics \cite{c89}. DEPTHS is able to decide on the concepts that should be added to the concept plan of a particular student along with a detailed lesson and a test plan for that concept. Each time the student performance significantly changes, the concept plan is created from scratch. The decision to add a new concept to the concept plan is made according to the curriculum sequence stored in the expert model and the performance of the student and his/her current knowledge stored in student model \cite{c89}.
			
			\subsubsection{Tutorial Dialog in Natural Language}
			Human tutors use conversational dialogs during tutoring  to deliver instructions. Early ITSs were not able to provide the use of natural language, discourse, or dialog based instruction. However, many modern ITSs use natural language \cite{c158}. The aim of this sub-section is to present how tutorial dialog techniques can be used to build interaction environments in ITSs along with some well-known dialog based ITSs found in the literature. 
			
			AutoTutor is a natural language tutoring system that has been developed for multiple domains such as computer literacy, physics, and critical thinking \cite{c138}. AutoTutor is a family of systems that has a long history. AutoTutor uses strategies of human tutors such as comprehension strategies, meta-cognitive strategies, self-regulated learning and meta-comprehension \cite{c160}\cite{c138}. In addition, AutoTutor incorporates learning strategies derived from learning research such as Socratic tutoring, scaffolding-fading, and frontier learning \cite{c158}. Benjamin et al. claim that the use of discourse in ITSs can facilitate new learning activities such as self-reflection, answering deep questions, generating questions and resolving conflicting statements \cite{c138}.
			
			In AutoTutor, the pedagogical interventions that occur between the system and students are categorized as positive feedback, neutral feedback, negative feedback, pump, prompt, hint, elaboration and splice/correction \cite{c159}. Latent Semantic Analysis (LSA) is used in AutoTutor as the backbone to represent computer literacy knowledge. The modules of AutoTutor are different from the traditional modules that have been identified and used in cognitive and constraint based tutors. The fact that AutoTutor uses language and discourse have led to the use of novel architectures (for more details on the architecture of AutoTutor see \cite{c159}). AutoTutor incorporates a variety computational architectures and learning methodologies, and has been shown to be very effective as a learning technology \cite{c138}.

			Atlas is an ITS that uses natural language dialogs to increase opportunities for students to construct their own knowledge \cite{c95}. The two main components of Atlas are APE \cite{c162}, the Atlas Planning Engine and CARMEL \cite{c161}, the natural language understanding component. APE is responsible for constructing and generating coherent dialogues while CARMEL understands and analyzes student's answers. Another conversational ITS is DeepTutor, an ITS developed for the domain of Newtonian physics \cite{c163}. A framework called learning progressions (LPs) used by the science education research community is integrated as a way to better model students' cognition and learning. The system implements conversational goals to accurately understand the student at each turn by analyzing the interaction that occurs between the system and student. Conversational goals such including coaching students to articulate expectations, correcting students' misconceptions, answering students' questions, feedback on students' contributions, and error handling \cite{c164}. In order to understand a student's contributions while interacting with DeepTutor, a semantic similarity task needs to be computed based on the quadratic assignment problem (QAP) \cite{c165}. An efficient branch and bound algorithm was developed to model QAP to reduce the explored space in search for the optimal solution.

			CIRCSIM-Tutor is a tutoring system in the area of cardiovascular physiology that incorporates natural
			language dialogue with the learner by using a collection of tutoring tactics that mimic expert human tutors \cite{c166} \cite{c167}. It can handle different syntactic constructions and lexical items such as sentence fragments
			and misspelled words. Tutoring tactics in CIRCSIM-Tutor are categorized into four major types as illustrated in Table \ref{table:1}. Theses evolved from the major types of tactics used in repeated pedagogical interventions: ask the next question, evaluate the user's response, recognize the user's answer, and if the answer is incorrect either provide a hint or the correct answer. The architecture of CIRCSIM-Tutor contains the following: a planner, a text generator, an input
			understander, a student model, a knowledge base, a problem solver and a screen manager \cite{c166}. CIRCSIM-Tutor showed significant improvement in students from pre-test to post-test. The input understander mechanism of the system was able to recognize and respond to over 95\% of students' inputs. Evens et al. suggest the use of APE for planning in tutoring sessions. 

			
			\begin{table}
				\renewcommand{\arraystretch}{1.3}
				\caption{Tutoring Tactics in CIRCSIM-Tutor Adopted From \cite{c166}}
				\label{table:1}
				\centering
				\begin{tabular}{|c||c|}
					\hline
					\textbf{Plan} & \textbf{Tactics\textbf{}} \\ 
					\hline
					Tutor       & Ask the student a series of questions.\\ 
					\hline
					Give answer & Ask the student to explain their answer. \\
					\hline
					Hint        & Remind the student (``Remember that....''). \\
					\hline
					Acknowledge & 4 possible cases (see below). \\
					\hline
				\end{tabular}

			\end{table} 	

Dialog based ITSs have the same main goal as traditional ITSs, which is to increase the level of engagement and learning gains. However, dialog based ITSs can use different dimensions of evaluation in classifying learner's responses, comprehending learner's contributions, modeling knowledge, and generating conversationally smooth tutorial dialogs. D'Mello and Graesser \cite{c168} conducted a study to describe how dialog based ITSs can be evaluated along these dimensions using AutoTutor as a case study. 	

\subsubsection{Spoken Dialogue}
			
			It is well-known that the best human tutors are more effective than the best computer tutors \cite{c169}. The main difference between human and computer tutors is the fact that human tutors predominantly use spoken natural language when interacting with learners. This raises the question of whether making the interaction more natural, such as by changing the modality of the computer tutor to spoken natural language dialogue, would decrease the advantage of human tutoring over computer tutoring. In fact, the majority of dialogue-based ITSs use typed student input. However, many potential advantages of using speech-to-speech interaction in the domain of ITSs have been found in the literature \cite{c169} \cite{c170}. One advantage is in terms of self-explanation, which gives the student a better opportunity to construct his/her knowledge \cite{c170}. For instance, Hauptmann et al. showed that self-explanation happens more often in speech than in typed interaction \cite{c171}. Another advantage is that speech interaction provides a more accurate student model. Students use meta-communication strategies such as hedges, pauses, and disfluencies, which allow the tutor to infer more information regarding student understanding. The following will discuses some computer tutors, which implement spoken dialogue \cite{c169} \cite{c170}.  
			
			ITSPOKE is an ITS which uses spoken dialogue for the purpose of providing spoken feedback and correcting misconceptions \cite{c173}. The student and the system interact with each other in English to discus the student's answers. ITSPOKE uses a microphone as an input device for the student's speech and sends the signal to the Sphinx2 recognizer \cite{c172}. Litman et al. showed that ITSPOKE is more effective than typed dialogue; however, there was no evidence that ITSPOKE increases student learning \cite{c169}. In addition, it was clear that speech recognition errors did not decrease learning.
			
			Another spoken ITS is SCoT (Spoken Conversational Tutor). SCoT's domain is shipboard damage control, which refers to  fires, flood and other critical situations that happen aboard Navy vessels \cite{c170}. Pon-Barry et al. suggested several challenges that ITS developers should be aware of when developing spoken language ITSs. First, repeated critical feedback from the tutor such as \textit{You made this mistake more than once} and\textit{ We discussed this same mistake earlier} cause negative effect. This suggests further work on better understanding and use of more tact in correcting user's misconceptions. Second, even though the accuracy of speech recognition is high, small recognition errors can make the tutor less effective.

			\section{Current Developments in ITS}
			In recent years, numerous effective and successful ITSs have been built. We have presented many such systems in prior sections of this paper. This section is intended to take a glance at a few significant recent systems and key areas of research focus at the current time.

			\subsection{Affective Tutoring System}
			Affective Tutoring Systems (ATS) are ITSs that can recognize human emotions (sad, happy, frustrated, motivated, etc.) in different ways \cite{c98}. It is important to incorporate the emotions of  students in the learning process because recent learning theories have established a link between emotions and learning, with the claim that cognition, motivation and emotion are the three components of learning \cite{c214}\cite{c99}. Over the last few years, there has been a great amount of interest in computing the learner's affective states in ITSs and studying how to respond to them in effective ways \cite{c101}. 
			
			Affective tutors use various techniques to enable computers to recognize, model, understand and respond to students' emotions in an effective manner. Knowing the emotional states of the student provides information on the student's psychological states and offers the possibility of responding appropriately \cite{c98}. A system can embed devices to detect a student's affective or emotional states. These include PC cameras, PC microphones, special mouses, and neuro-headsets among others. These devices are responsible for identifying physical signals such as facial image, voice, mouse pressure, heart rate and stress level. These signals are then sent to the system to be processed. Consequently, the emotional state is obtained in real time. The ATS objective is to change a negative emotional state (e.g., confused) to a positive emotional state (e.g. committed) \cite{c100}. 
			
			In \cite{c102}, the learners' affective states are detected by monitoring their gross body language (body position and arousal) as they interact with the system. An automated body pressure measurement system is also used to capture the learner's pressure. The system detects six affective states of the learner: confusion, flow, delight, surprise, boredom and neutral. If the system realizes that the student is bored, the tutor stimulates him by presenting engaging tasks. If frustration is detected, the tutor offers encouraging statements or corrects information that the learner is experiencing difficulty with. Experiments suggest that that boredom and flow might best be detected from body language although the face plays a significant role in conveying confusion and delight.
			
			Jraidi et al. present an ITS that acts differently when the student is frustrated  \cite{c103}. For example, it may provide problems similar to ones in which the student has been successful to help the student. In case of boredom, the system provides an easier problem to motivate the student again or provides a more difficult problem if the problem seems too easy. Another approach used in the system to respond to student emotions integrates a virtual pedagogical agent called a \textit{learning companion} to allow affective real time interaction with the learners. This agent can communicate with the learner as a study partner when solving problems, or provide encouragement and congratulatory messages, appearing to care about the learner. In other words, these agents can provide empathic responses which mirror the learner's emotional states \cite{c100}.
			
			Wolf and his colleagues also implement an empathetic learning companion that reflects the last expressed emotion of the learner as long as the emotion is not negative such as frustration or boredom \cite{c104}\cite{c105}. The companion responds in full sentences providing feedback with voice and emotion. The presence of someone who appears to care can be motivating to the learners. Studies show that students who use the learning companion increased their math understanding and level of interest, and show reduced boredom. Another affective tutoring system that uses an empathetic companion to respond to learner emotion is a system that practices interview questions with  users \cite{c106}. The system perceives the user's emotion by measuring skin conductance and then takes appropriate actions. For instance, the agent displays concern for a user who is aroused and has a negatively valenced emotion, e.g., by saying ``I am sorry that you seem to feel a bit bad about that question''. Their study shows that users receiving feedback with empathy are less stressed when asked interview questions. 
			
			\subsection{Cultural Awareness in Education }
			In recent years, special attention is being paid to the issues that arise in the context of delivering  education in a globalized society \cite{c204}.  Researchers in the field of ITS and learning technologies are increasingly concerned about how learning technology systems can be adapted across a diversity of cultures. Nye in 2015 \cite{c205} addressed the barriers faced by ITSs entering the developing world. Barriers such as lack of student computing skills, problems arising due to multiplicity of languages and cultures, etc., were presented along with existing solutions. An analysis of  student help seeking behaviors in ITSs across different cultures was conducted by Ogan et al. \cite{c206}.
			
			 Models of help seeking behaviors during learning have been developed based on datasets of students in three different countries: Costa Rica, the Philippines, and the United States. Ogan et al. find that help seeking behaviors across different cultures is not substantially transferable. This finding suggests the need to replicate research to understand student behaviors. Mohammed and Mohan \cite{c207} take the first step toward tackling this issue. Their system provides learners with some control over their cultural preferences including problem description, feedback, and presentation of images and hints. Deployment of such systems has provided researchers with the opportunity to experimentally investigate phenomena surrounding the social acceptability of non-dominant language use in education, and its effects on learning. 
			
			\subsection{Game-based Tutoring Systems}
			The novelty of an ITS and its interactive components is quite engaging when they are used for short periods of time (e.g., hours), but can be monotonous and even annoying when a student is required to interact with an ITS for weeks or months \cite{c107}. The underlying idea for game based learning is that students learn better when they are having fun and engaged in the learning process.
			
			Game based tutoring systems engage learners to interact actively with the system, thereby making them more motivated to use the system for a longer time \cite{c108}. Whereas the ITS principles maximize learning, the game technologies maximize motivation. Instead of learning a subject in a conventional and traditional way, the students play an educational game which successfully integrates game strategies with curriculum-based contents. Although there is no overwhelming evidence supporting the effectiveness of educational game based systems over computer tutors, it has been found that educational games have advantages over traditional tutoring approaches \cite{c109}\cite{c110}. Moreno and Mayer \cite{c111} summarize characteristics of educational games that make them enjoyable to operate. These are interactivity, reflection, feedback, and guidance.

\begin{figure}[!t]
				\centering
				\includegraphics[width=2.8in]{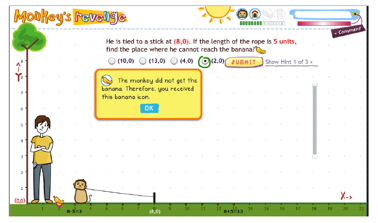}
				\caption{Math-learning Environment with Game-like Elements \cite{c112}}
				\label{fig6}
			\end{figure}

			To enhance both engagement and learning, Rai and Beck implemented game-like elements in their math tutor \cite{c112}. The system provides a math learning environment and the students engage in a narrated visual story. Students help story characters solve the problem in order to move the story forward as shown in Figure \ref{fig6}. Students receive feedback and bug messages as when using a traditional tutor. The study found that students are more likely to interact with the version of the math tutor that contains game-like elements; however, the authors suggest adding more tutorial features to a game-like environment for higher levels of learning.

			Another tutoring system that uses an educational game approach is Writing Pal (W-Pal), which is designed to help students across multiple phases of the writing process \cite{c113}. Crystal Island is a narrative-centered learning environment in biology, where students attempt to discover the identity and source of an infectious disease on a remote island. The student (player) is involved in a scenario of meeting a patient and attempts to perform a diagnosis. The study of educational impact using a game based system by Lester at el. \cite{c114} found that students answer more questions correctly on the post-test than the pre-test, and this finding was statistically significant. Additionally, there was a strong relationship between learning outcomes, in-game problem solving and increased engagement \cite{c114}.
			
			\subsection{Adaptive Intelligent Web Based Educational System (AIWBES)}
			Adaptive Intelligent Web Based Educational Systems (AIWBES) or adaptive hypermedia provide an alternative to the traditional, just-put-it-on-the-web approach in the development of web based educational courseware. An AIWBES provides adaptivity in terms of goals, preferences, and knowledge of individual students during interaction with the system \cite{c115}.

			The area of ITSs inspired early research on adaptive educational hypermedia, which combine ITSs and educational hypermedia. During the development of the early ITSs, the concern was to support students in solving problems and how to overcome the lack of learning material. The required knowledge was acquired by developers attending lectures or reading textbooks. As computers became more powerful, ITS researchers integrated ITS features with the learning material. Many research groups have found that combining hypermedia systems with an ITS can lead to more functionality than traditional static educational hypermedia \cite{c117}.
			
			A number of systems have been developed under the category of AIWBES. ELM-ART (ELM Adaptive Remote Tutor) is a WWW based ITS to support learning programming in Lisp. It has been used in distance learning to not only support course material from the textbook, but also to provide problem solving support. Adaptive navigation through the material was implemented to support learning by individual students. The system classifies the content of a page to be as ready to be learned or not ready to be learned  because some prerequisite knowledge has not been learned yet \cite {c120}. In addition, the links are sorted depending on the relevancy to the current student state so the students know which are the most similar situations or most relevant web pages. When the student enters a page which contains a chunk of prerequisite knowledge to be learned, the system alerts the student about the prerequisite and suggests additional links to textbook and manual pages regarding them. In case the student struggles with understanding some contents or solving a problem,  he/she can use the help button \cite {c120}. Empirical studies have shown that hypermedia systems in conjunction with tutoring tools can be helpful with a self-learner \cite{c121}. Other adaptive intelligent hypermedia systems that have been used by hundreds of students include AHA! \cite{c147} and InterBook \cite{c148}, which have been shown to help student learn fast and better \cite{c122}.

			\subsection{Collaborative Learning}
			Current educational research suggests collaborative learning or group-based learning increases the learning performance of a group as well as individual learning outcomes \cite{c130} \cite{c213}. In a collaborative learning environment, students learn in groups via interactions with each other by asking questions, explaining and justifying their opinions, explaining their reasoning, and presenting their knowledge \cite {c131}. A number of researchers have pointed out the importance of a group learning environment and how significantly effective it is in term of learning gain \cite{c132}.
			
			Recently, there has been a rise in interest in implementing collaborative learning in tutoring systems to show the benefits obtained from interactions among students during problem solving. Kumar and Rose, in 2011, built intelligent interactive tutoring systems CycleTalk and WrenchTalk that support collaborative learning environments in the engineering domain \cite{c13}. Teams of two or more students work on the same task when solving a problem. They conducted a number of experiments to investigate the effectiveness of collaborative learning and how to engage the students more deeply in instructional conversations with the tutors using teaching techniques such as Attention Grabbing, Ask when Ready and Social Interaction Strategies. It was found that students who worked in pairs learned better than students who worked individually \cite {c133}\cite{c134}.  Another tutoring system that supports collaborative learning is described in \cite{c135} for teaching mathematical fractions. 
			
			\subsection{Data Mining in ITSs}	
				Data mining or knowledge discovery in databases as it is alternatively called, is the process of analyzing large amounts of data for the purpose of extracting and discovering useful information \cite{c174}. Data mining has been used in the field of ITSs for many different purposes. For instance, it has been used to identify learners who game the system. Gaming the system or off-task behavior which is defined as ``attempting to succeed in the environment by exploiting properties of the system rather than by learning the material and trying to use that knowledge to answer correctly'' \cite{c179}. Identifying situations where the system has been gamed has been the focus for many researchers in recent years. Additional discussions on mining student datasets can be found in \cite{c182} \cite{c183}.
			
			Another use of data mining in ITSs is to detect student affect. Detecting student's affective states can potentially increase the engagement level and learning outcomes as stated by Baker et al. \cite{c183}. For example, classification methods have been used in automating detectors to predict student states, including boredom, engaged concentration, frustration, and confusion \cite{c184}. Similarly, classification methods have been used to detect affect such as joy and distress \cite{c185}.
			
			Another use of data mining is automatically discovering a partial problem space from logged user interactions rather than traditional techniques where domain experts have to provide the source of the knowledge. As an example, clustering methods including sequential pattern mining \cite{c186} and association rule discovery \cite{c187} are used in RomanTutor \cite{c188} to extract problem space and support tutoring services \cite{c189}. Interested readers are referred to read \cite{c175} for more details.

			\subsection{Authoring Tools}
			ITS researcher teams have been interested in simplifying the building process for ITSs by making authoring of ITSs more accessible and affordable to designers and teachers. Authoring tools in the domain of ITSs can be categorized  in different dimensions such as tools that require programming skills and those that do not, pedagogy-oriented and performance-oriented \cite{c190}, or paradigm specific such as model tracing system, and constraint based tutor \cite{c191}. 
			
			SModel \cite{c192} and Tex-Sys \cite{c193} are tools that fall into the category of tools that require programming skills. SModel is a Bayesian student modeling component, which provides services to a group of agents in the CORBA platform. Tex-Sys is another example in the same category, which provides a generic module for designing ITS components (domain model, and student model) for any given domain. 
			
			Examples of authoring tools categorized as pedagogy-oriented are REDEEM \cite{c194} and CREAM-Tools \cite{c195}. Pedagogy-oriented tools are those that concentrate on how to deliver and sequence a package of content \cite{c191}. REDEEM provides reusability of existing domain material and then provides authoring tactics on how to teach this material, tactics such as sequencing of contents and learning activities. Similarly, CREAM-Tools focuses on the operations required to develop  curriculum content, taking into account aspects of the domain, and pedagogy and didactic requirements. Performance-oriented tools are those that concentrate on the learner's performance to provide a rich environment of skills for the learners to practice and receive system responses \cite{c191}. Examples of authoring tools that belong to this category are Demonstr8 \cite{c200}, XAIDA \cite{c201} and Knowledge Construction Dialog (KCD) \cite{c202}.
			
			In recent years, there has been a great interest in building authoring tools that are specific to certain paradigms and do not require programing skills in order to allow for sharing of components across ITSs and reduce development costs \cite{c191}. Cognitive Tutor Authoring Tools (CTAT) \cite{c196} provides a set of authoring tool specific for model tracing tutors and example tracing tutors \cite {c27}. CTAT provides step by step guidance for problem solving activities as well as how to adaptively select problems based on a Bayesian student model. Authoring Software Platform for Intelligent Resources in Education (ASPIRE) is also a paradigm specific authoring tool for constraint based models \cite{c197}. ASPIRE supports authoring the domain model, enabling the subject experts to easily develop constraint based tutors. Another authoring tool that falls into this category is AutoTutor Script Authoring Tool (ASAT) \cite{c198}. ASAT  facilitates developing components of AutoTutor, integrating conversations into learning systems. Finally, the Generalized Intelligent Framework for Tutoring (GIFT) is a framework and set of tools for developing intelligent and adaptive tutoring systems \cite{c199}. GIFT supports a variety of services that include domain knowledge representation, performance assessment, course flow, pedagogical model and student model. 
			
			\section{Discussion} 
			
			ITSs are educational systems that attempt to adapt to special needs of individual learners. What makes ITSs different from other educational systems are their abilities to keep track of cognitive states of individual students and respond appropriately. ITSs have received ample attention in  disciplines such as cognitive science, education and computer science. The ultimate goal is to achieve the possibility of mimicking expert human tutors in the way they teach and interact with learners. The following paragraphs present some shortcomings of ITSs from the authors' point of view.
			   
			People with special needs generally suffer from slower learning pace; therefore, special attention should be paid to investigate how an ITS can be specialized to improve their learning skills, say reading and writing skills. ITSs have already proven their pedagogical effectiveness and helped improve learners' outcomes. Consequently, ITS systems are likely to be helpful to either adults or children with special needs in their quest to achieve their learning goals. Obviously, one must incorporate proven successful strategies for teaching such individuals into the models as an ITS is constructed. Effective and targeted ITSs or ITS modules are likely to be of great assistance in teaching individuals with cognitive disabilities such as Down Syndrome, traumatic brain injury, or dementia, as well as for less severe cognitive conditions such as dyslexia, attention deficit disorder and dyscalculia.  
			
			Data mining in the context of ITSs has drawn a significant attention recently, since the findings can be used to elaborate  learning outcomes in many ways, as discussed in the previous section. It will be interesting and beneficial if ITS research has access to long-term data available to follow a cohort of students for many years.  For example, it may be worthwhile to track students who benefit from their interactions with ITSs during middle or high school (early education stage) until later education stages and beyond. Researchers should keep track from an early age until the time of graduation from college including major, Grade Point Average (GPA), and other variables. However, this is likely to be prohibitively expensive. 
			
			For instance, students interacting with an ITS in the mathematics domain during early stages of education can be tracked until their college graduation. This data can further be analyzed in-depth to find relations if any, between the ITS data and the learning outcomes in math related majors years later, considering  GPA as a measure of proficiency. Based on the result of analysis, we may be able to suggest to students in the future if math related majors are appropriate disciplines for them to pursue. To the best of our knowledge, there is not yet any such datasets. Several benefits can be gained from this direction of research. First, there are hundreds of ITSs in a variety of subject domains. Thus, it is possible that the findings may be extended to other disciplines as well. Second, this direction of research may also increase the popularity of ITSs as a new educational tool for the purpose of assisting students' decisions regarding majors to pursue. Finally, it may also, provide us enough time to significantly improve students' skills and make them ready for their graduation major basic skills.

			\section{Conclusion}
			The gap between human tutors and software tutors in the form of ITSs is narrowing, but not closed, even remotely. Many different models exist for representing knowledge, teaching styles and student knowledge. Each model has its benefits and shortcomings. Hybrid models have also been created to enhance and strengthen traditional models. Even with the many unanswered questions that continue to surround the principles of human thought and learning, many ITSs have been implemented and tested. ITSs show  promise in possible standardizing  and implementing aspects of human learning, but still have many limitations to overcome. The close marriage of ITSs with  AI and psychology shows continued promise for the advancement of ITSs. While there are no ITSs to date that possess the cognitive awareness of an actual human tutor, the availability, readiness, and consistency of ITSs may make them a competitive alternative to human tutors in the future where cost, time, and scale are the friends of the ITS.




\end{document}